\documentclass
[superscriptaddress,secnumarabic,amssymb,amsmath,nobibnotes,aps,prd,showkeys,showpacs,nofootinbib,onecolumn,12pt]{revtex4}%
\usepackage{graphicx}
\usepackage{subfigure}
\usepackage{epsf}
\usepackage{bm}
\usepackage{amsmath}
\usepackage{amsfonts}
\usepackage{amssymb}%
\providecommand{\U}[1]{\protect\rule{.1in}{.1in}}

\newcommand{\be}{\begin{equation}}
\newcommand{\ee}{\end{equation}}

\newcommand{\mincir}{\raise
-3.truept\hbox{\rlap{\hbox{$\sim$}}\raise4.truept\hbox{$<$}\ }}
\newcommand{\magcir}{\raise
-3.truept\hbox{\rlap{\hbox{$\sim$}}\raise4.truept\hbox{$>$}\ }}

\begin{document}
\title{Dynamical analysis and cosmological evolution in Weyl integrable gravity}
\author{Andronikos Paliathanasis}
\email{anpaliat@phys.uoa.gr}
\affiliation{Institute of Systems Science, Durban University of Technology, Durban 4000,
South Africa}
\affiliation{Instituto de Ciencias F\'{\i}sicas y Matem\'{a}ticas, Universidad Austral de
Chile, Valdivia 5090000, Chile}

\begin{abstract}
We investigate the cosmological evolution for the physical parameters in Weyl
integrable gravity in a Friedmann--Lema\^{\i}tre--Robertson--Walker universe
with zero spatially curvature. For the matter component, we assume that it is
an ideal gas, and of the Chaplygin gas. From the Weyl integrable gravity a
scalar field is introduced by a geometric approach which provides an
interaction with the matter component. We calculate the stationary points for
the field equations and we study their stability properties. Furthermore, we
solve the inverse problem for the case of an ideal gas and prove that the
gravitational field equations can follow from the variation of a Lagrangian
function. Finally, variational symmetries are applied for the construction of
analytic and exact solutions.

\end{abstract}
\date{\today}
\maketitle

\section{Introduction}

\label{sec1}

The cosmological constant component in the Einstein-Hilbert Action Integral,
is the simplest dark energy candidate to describe of the recent acceleration
phase of the universe, as it is provided by the cosmological observations
\cite{pl1}. In the so-called $\Lambda$CDM cosmology the universe is considered
to be homogeneous and isotropic, described by the
Friedmann--Lema\^{\i}tre--Robertson--Walker (FLRW) geometry with spatially
flat term, where the matter component consists of the cosmological constant
and a pressureless fluid source which attributes the dark matter component of
the universe. The gravitational field equations are of second-order and can be
integrated explicitly. Indeed, the field equations can be reduced to that of
the one-dimensional \textquotedblleft hyperbolic oscillator\textquotedblright.
However, as the cosmological observations are improved, $\Lambda$-cosmology
loses the important position in the \textquotedblleft
armoury\textquotedblright\ of cosmologists. For an interesting discussion on
the subject we refer the reader to the recent review \cite{leon1}.
Furthermore, because of the simplicity of the field equations in $\Lambda
$-cosmology, the cosmological constant term cannot provide a solution for the
description of the complete cosmological evolution and history.

In order to solve these problems, cosmologists have introduced various
solutions in the literature by introducing new degrees of freedom in the field
equations. Time-varying $\Lambda$ term, scalar fields and fluids with
time-varying equation of state parameters, like the Chaplygin gases have been
proposed to modify the energy-momentum tensor of the field equations
\cite{sf1,sf2,sf3,sf4,sf5,sf5a}. On the other hand, a different approach is
inspired by the modification of the Einstein-Hilbert Action integral, and
leads to the family of theories known as alternative/modified theories of
gravity \cite{sf6,sf7,sf8}. Another interesting consideration is the
interaction between the various components consisted of the energy momentum
tensor \cite{in0}. Interaction in the dark components of the cosmological
model, that is, between, the dark energy and the dark matter terms is
supported by cosmological observations \cite{in2,in4,in5,in6}.

For a given proposed dark energy mode model, there are systematic methods for
the investigation of the physical properties of the model. The derivation of
exact and analytic solutions is an essential approach because analytic
techniques can be used for the investigation of the cosmological viability of
the model \cite{as1,as2,as3,as4}. Furthermore, from the analysis of the
asymptotic dynamics, that is, of the determination of the stationary points,
the complete cosmological history can be constructed \cite{dn1,dn2,dn3}.
Indeed, constraints for the free parameters of a given model can be
constructed through the analysis of the stationary points and the specific
requirements for the stability of the stationary points
\cite{dn4,dn5,dn6,dn7,dyn8}.

In this piece of work, we study the evolution of the cosmological dynamics for
the theory known as Weyl integrable gravity (WIG)
\cite{sc1,sc2,sc3,sc4,sc5,sc6,sc7}. In WIG a scalar field is introduced into
the Einstein-Hilbert Action Integral by a geometric construction approach.
Indeed, in Riemannian geometry the basic geometric object is the covariant
derivative~$\nabla_{\mu}$ and the metric tensor $g_{\mu\nu},~$such that it has
no metricity component,i.e. $\nabla_{\kappa}g_{\mu\nu}=0~$\cite{salim96}. In
Weyl geometry the fundamental geometric objects are the gauge vector field
$\omega_{\mu}$ and the the metric tensor $g_{\mu\nu}$, such that
$\tilde{\nabla}_{\kappa}g_{\mu\nu}=\omega_{\kappa}g_{\mu\nu}$, where now
$\tilde{\nabla}_{\mu}$ notes the covariant derivative with respecto the affine
connection $\tilde{\Gamma}_{\mu\nu}^{\kappa}~$\ which is defined as
$\tilde{\Gamma}_{\mu\nu}^{\kappa}=\Gamma_{\mu\nu}^{\kappa}-\omega_{(\mu}%
\delta_{\nu)}^{\kappa}+\frac{1}{2}\omega^{\kappa}g_{\mu\nu}$. \ When
$\omega_{\mu}$ is defined by a scalar field $\phi$, $\tilde{\Gamma}_{\mu\nu
}^{\kappa}$ describes the affine connection for the conformal metric
$\tilde{g}_{\mu\nu}=\phi g_{\mu\nu}$. The field equations of the WIG in the
vacuum are equivalent to that of General Relativity with a massless scalar
field, with positive or negative energy density. However, when a matter source
is introduced, interaction terms appear as a natural consequence of the
geometry of the theory \cite{salim96}. In geometric terms of interaction
context , we investigate the dynamics of the cosmological field equations so
that we construct the cosmological history and investigate the viability of
the theory. Furthermore, the integrability property for the field equations is
investigated by using the method of variational symmetries for the
determination of conservation laws.

In Section \ref{sec2} we present the basic elements for the WIG theory.
Furthermore, we write the field equations for our cosmological model in a
spatially flat FLRW background space. In Section \ref{sec3} we present the
main results of our analysis in which we discuss the asymptotic dynamics for
the field equations in the cases for which the matter source is an ideal gas,
or a Chaplygin gas. Moreover, we investigate the dynamics in the presence of
the cosmological constant term. In Section \ref{sec4} we show that the field
equations have a minisuperspace description when the matter source is an ideal
gas. Specifically, we solve the inverse problem and we construct a point-like
Lagrangian which describes the cosmological field equations. With the use of
the variational symmetries we determine a conservation law and we present the
analytic solution for the field equations by using the Hamilton-Jacobi
approach. Our results are summarized in Section \ref{sec5}.

\section{Weyl Integrable gravity}

\label{sec2}

Consider the two conformal related metric tensors $g_{\mu\nu}$, $\tilde
{g}_{\mu\nu}$ such that $\tilde{g}_{\mu\nu}=\phi g_{\mu\nu}$. The Christoffel
symbols of the two conformal related metrics are related as
\begin{equation}
\tilde{\Gamma}_{\mu\nu}^{\kappa}=\Gamma_{\mu\nu}^{\kappa}-\phi_{,(\mu}%
\delta_{\nu)}^{\kappa}+\frac{1}{2}\phi^{,\kappa}g_{\mu\nu}.
\end{equation}

In Weyl geometry the fundamental objects are the metric tensor $g_{\mu\nu}$
and the covariant derivative $\tilde{\nabla}_{\mu}$ defined by the Christoffel
symbols $\tilde{\Gamma}_{\mu\nu}^{\kappa}$. Hence, the curvature tensor is
defined
\begin{equation}
\tilde{\nabla}_{\nu}\left(  \tilde{\nabla}_{\mu}u_{\kappa}\right)
-\tilde{\nabla}_{\mu}\left(  \tilde{\nabla}_{\nu}u_{\kappa}\right)  =\tilde
{R}_{\kappa\lambda\mu\nu}u^{\lambda}. \label{ww.03}%
\end{equation}
Consequently, the Ricci tensors of the two conformal metrics are related as
follow%
\begin{equation}
\tilde{R}_{\mu\nu}=R_{\mu\nu}-\tilde{\nabla}_{\nu}\left(  \tilde{\nabla}_{\mu
}\phi\right)  -\frac{1}{2}\left(  \tilde{\nabla}_{\mu}\phi\right)  \left(
\tilde{\nabla}_{\nu}\phi\right)  -\frac{1}{2}g_{\mu\nu}\left(  \frac{1}%
{\sqrt{-g}}\tilde{\nabla}_{\nu}\tilde{\nabla}_{\mu}\left(  g^{\mu\nu}\sqrt
{-g}\phi\right)  -g^{\mu\nu}\left(  \tilde{\nabla}_{\mu}\phi\right)  \left(
\tilde{\nabla}_{\nu}\phi\right)  \right)  , \label{ww.04}%
\end{equation}
thus the Ricci scalar%
\begin{equation}
\tilde{R}=R-\frac{3}{\sqrt{-g}}\tilde{\nabla}_{\nu}\tilde{\nabla}_{\mu}\left(
g^{\mu\nu}\sqrt{-g}\phi\right)  +\frac{3}{2}\left(  \tilde{\nabla}_{\mu}%
\phi\right)  \left(  \tilde{\nabla}_{\nu}\phi\right)  . \label{ww.05}%
\end{equation}

In WIG the the foundamental Action Integral is defined by using the Weyl Ricci
scalar $\tilde{R}$ and the scalar field $\phi$ by the expression
\begin{equation}
S_{W}=\int dx^{4}\sqrt{-g}\left(  \tilde{R}+\xi\left(  \tilde{\nabla}_{\nu
}\left(  \tilde{\nabla}_{\mu}\phi\right)  \right)  g^{\mu\nu}-\Lambda\right)
, \label{ww.06}%
\end{equation}
where $\xi$ is a coupling constant. From (\ref{ww.06}) we observe that $\phi$
is a massless scalar field. However, in a more general consideration a
potential function may be considered.

From the Action Integral (\ref{ww.06}) the Weyl-Einstein equations are as
\cite{salim96}
\begin{equation}
\tilde{G}_{\mu\nu}+\tilde{\nabla}_{\nu}\left(  \tilde{\nabla}_{\mu}%
\phi\right)  -\left(  2\xi-1\right)  \left(  \tilde{\nabla}_{\mu}\phi\right)
\left(  \tilde{\nabla}_{\nu}\phi\right)  +\xi g_{\mu\nu}g^{\kappa\lambda
}\left(  \tilde{\nabla}_{\kappa}\phi\right)  \left(  \tilde{\nabla}_{\lambda
}\phi\right)  -\Lambda g_{\mu\nu}=0, \label{ww.08}%
\end{equation}
where $\tilde{G}_{\mu\nu}$ is the Weyl Einstein tensor. By using the
Riemannian Einstein tensor $G_{\mu\nu},$ the Weyl-Einstein field equations
(\ref{ww.08}) become \cite{salim96}%
\begin{equation}
G_{\mu\nu}-\lambda\left(  \phi_{,\mu}\phi_{,\nu}-\frac{1}{2}g_{\mu\nu}%
\phi^{,\kappa}\phi_{,\kappa}\right)  -\Lambda g_{\mu\nu}=0, \label{ww.11}%
\end{equation}
where $\lambda$ is defined as $2\lambda\equiv4\xi-3$. Equations (\ref{ww.11})
are nothing else than the field equations of Einstein's General Relativity
with a massless scalar field. When $\lambda>0$, the scalar field $\phi$ is a
quintessence while, when $\lambda<0$, $\phi$ is a phantom field \cite{salim96}

Moreover, for the equation of motion of the scalar field $\phi$, the
Klein-Gordon equation is \cite{salim96}%
\begin{equation}
\left(  \tilde{\nabla}_{\nu}\left(  \tilde{\nabla}_{\mu}\phi\right)  \right)
g^{\mu\nu}+2g^{\mu\nu}\left(  \tilde{\nabla}_{\mu}\phi\right)  \left(
\tilde{\nabla}_{\nu}\phi\right)  =0, \label{ww.09}%
\end{equation}
or by using the Riemannian covariant derivative $\nabla_{\mu}$, expression
(\ref{ww.09}) is written in the usual form $g^{\mu\nu}\nabla_{\nu}\nabla_{\mu
}\phi=0.$

As it was found in \cite{salim96}, the introduction of a perfect fluid in the
gravitational model leads to the following set of gravitational field
equations \cite{salim96}%
\begin{equation}
\tilde{G}_{\mu\nu}+\tilde{\nabla}_{\nu}\left(  \tilde{\nabla}_{\mu}%
\phi\right)  -\left(  2\xi-1\right)  \left(  \tilde{\nabla}_{\mu}\phi\right)
\left(  \tilde{\nabla}_{\nu}\phi\right)  +\xi g_{\mu\nu}g^{\kappa\lambda
}\left(  \tilde{\nabla}_{\kappa}\phi\right)  \left(  \tilde{\nabla}_{\lambda
}\phi\right)  -\Lambda g_{\mu\nu}=e^{-\frac{\phi}{2}}T_{\mu\nu}^{\left(
m\right)  }, \label{ww.12}%
\end{equation}
that is,%
\begin{equation}
G_{\mu\nu}-\lambda\left(  \phi_{,\mu}\phi_{,\nu}-\frac{1}{2}g_{\mu\nu}%
\phi^{,\kappa}\phi_{,\kappa}\right)  -\Lambda g_{\mu\nu}=e^{-\frac{\phi}{2}%
}T_{\mu\nu}^{\left(  m\right)  }, \label{ww.14}%
\end{equation}
where $T_{\mu\nu}^{\left(  m\right)  }=\left(  \rho_{m}+p_{m}\right)  u_{\mu
}u_{\nu}+p_{m}g_{\mu\nu}$.

Moreover, the modified Klein-Gordon equation follows \cite{salim96}%
\begin{equation}
-g^{\mu\nu}\nabla_{\nu}\nabla_{\mu}\phi=\frac{1}{2\lambda}e^{-\frac{\phi}{2}%
}\rho_{m}. \label{ww.15}%
\end{equation}

Equation (\ref{ww.15}) follows from the identity $G_{;\nu}^{\mu\nu}=0$, which
provides the conserve of the effective energy-momentum tensor.

\subsection{FLRW spacetime}

Following the cosmological principle, in very large scales the universe is
considered to be isotropic and homogeneous. Hence, the physical space is
described by the FLRW spacetime, where the three-dimensional surface is a
maximally symmetric space and admits six isometries. However, from
cosmological observations the spatial curvature is very small, which means
that we can consider as background space the spatially flat FLRW metric%
\begin{equation}
ds^{2}=-dt^{2}+a^{2}\left(  t\right)  \left(  dr^{2}+r^{2}\left(  d\theta
^{2}+\sin^{2}\theta d\varphi^{2}\right)  \right)  . \label{ww.16}%
\end{equation}

Moreover, we assume the comoving observer $u_{\mu}=\delta_{\mu}^{t}$, with
expansion rate $\theta=3\frac{\dot{a}}{a}$, \ for the line element
(\ref{ww.16}) and for a scalar field $\phi=\phi\left(  t\right)  $, the
gravitational field equations are
\begin{equation}
\frac{\theta^{2}}{3}-\frac{\lambda}{2}\dot{\phi}^{2}-\Lambda-e^{-\frac{\phi
}{2}}\rho_{m}=0, \label{ww.17}%
\end{equation}%
\begin{equation}
\dot{\theta}+\frac{1}{3}\theta^{2}+\frac{1}{2}e^{-\frac{\phi}{2}}\left(
\rho_{m}+3p_{m}\right)  +\lambda\dot{\phi}^{2}-\Lambda=0, \label{ww.18}%
\end{equation}%
\begin{equation}
\ddot{\phi}+\theta\dot{\phi}+\frac{1}{2\lambda}e^{-\frac{\phi}{2}}\rho_{m}=0
\label{ww.19a}%
\end{equation}
and
\begin{equation}
\dot{\rho}_{m}+\theta\left(  \rho_{m}+p_{m}\right)  -\rho_{m}\dot{\phi}=0.
\label{ww.20a}%
\end{equation}

From the modified Friedmann equations we observe the existence of a nonzero
interacting term for scalar field $\phi$ and the matter component $\rho_{m}$.
When $\lambda>0$, energy decays from scalar field to the $\rho_{m}$, while for
$\lambda<0$ energy decays from $\rho_{m}$ to the field $\phi$. Furthermore,
the effective equation of state parameter for the effective cosmological
matter is defined as $w_{eff}=-1-2\frac{\dot{\theta}}{\theta^{2}}$.

Finally, for the nature of the matter source $\rho_{m}$ in the following we
consider that $\rho_{m}$ is an ideal gas, or a Chaplygin gas.

\section{Cosmological dynamics}

\label{sec3}

We continue our analysis with the investigation of the stationary points for
the cosmological field equations. In order to proceed with the study we define
the new dimensionless variables in the context of $\theta$-normalization%
\begin{equation}
x=\sqrt{\frac{3}{2}}\frac{\dot{\phi}}{\theta}~,~\Omega_{\Lambda}%
=\frac{3\Lambda}{\theta^{2}}~,~\Omega_{m}=\frac{3\rho_{m}}{\theta^{2}%
}e^{-\frac{\phi}{2}}~ \label{ww.25}%
\end{equation}
where for the equation of state parameter for the matter source we consider
(i) ideal gas $p_{m}=\left(  \gamma-1\right)  \rho_{m}$,~$0\leq\gamma<2$, and
(ii) Chaplygin gas $p_{m}=\frac{A_{0}}{\rho_{m}^{\alpha}}$, $\alpha\geq1$.
Moreover, we define the new independent parameter to be $\tau=\ln\left(
a\right)  $, such that $x^{\prime}=\frac{dx}{d\tau}$.

At the stationary points the effective equation of the state parameter is
defined as $w_{eff}=w_{eff}\left(  x,\Omega_{\Lambda},\Omega_{m}\right)  $, so
that the asymptotic solution is described by the scale factor $a\left(
t\right)  =a_{0}t^{\frac{2}{3\left(  1+w_{eff}\right)  }}$, $w_{eff}\neq-1$
and $a\left(  t\right)  =a_{0}e^{H_{0}t}$, when $w_{eff}=-1$.

\subsection{Ideal gas with $\Lambda=0$}

Assume the equation of state of an ideal gas $p_{m}=\left(  \gamma-1\right)
\rho_{m}$, without the cosmological constant term. Then in the new
dimensionless variables\ (\ref{ww.25}) the field equations are%
\begin{equation}
\Omega_{m}=1-\lambda x^{2}~, \label{ww.31}%
\end{equation}%
\begin{equation}
x^{\prime}=-\frac{\left(  1-\lambda x^{2}\right)  \left(  \sqrt{6}-6\left(
\gamma-2\right)  \lambda x\right)  }{12\lambda}\text{.} \label{ww.32}%
\end{equation}
Moreover, $\Omega_{m}$ is bounded as $0\leq\Omega_{m}\leq1$, such that the
solution is physically acceptable, that is, from (\ref{ww.31}) it follows that
there are physical stationary points only when $\lambda>0$.

The stationary points of equation (\ref{ww.32}) are
\begin{equation}
A_{1}^{\pm}:x_{1}^{\pm}=\frac{1}{\sqrt{\lambda}}~,~A_{2}:x_{2}=\frac{1}%
{\sqrt{6}\left(  \gamma-2\right)  \lambda}.
\end{equation}
Points $x_{1}^{\pm}$ describe asymptotic solutions where only the scalar field
contributes to the cosmological fluid. The effective equation of state
parameter is derived to be $w_{eff}\left(  x_{1}^{\pm}\right)  =1$, from which
we infer that the solution is that of a stiff fluid. On the other hand, the
point $x_{2}$ is physically acceptable when $\lambda\geq\frac{1}{6\left(
\gamma-2\right)  ^{2}}$, and the point describes a scaling solution with
$w_{eff}\left(  x_{2}\right)  =-1+\gamma+\frac{1}{6\lambda\left(
\gamma-2\right)  }$. For$~\gamma<\frac{2}{3}$, $\lambda>\frac{1}{8\left(
1-2\gamma\right)  +6\gamma^{2}}$ it follows that $w_{eff}\left(  x_{2}\right)
<-\frac{1}{3}$ which means that the asymptotic solution describes an
accelerated universe, where in the limit $\lambda=\frac{1}{8\left(
1-2\gamma\right)  +6\gamma^{2}}$, the asymptotic solution is that of the de
Sitter universe.

We proceed with the investigation of the stability properties for the
stationary points. We linearize equation (\ref{ww.32}) and we find the
eigenvalues $e_{1}\left(  x_{1}^{\pm}\right)  =2-\gamma\mp\frac{1}%
{\sqrt{6\lambda}}$, $e_{1}\left(  x_{2}\right)  =-1+\frac{\gamma}{2}+\frac
{1}{12\lambda\left(  2-\gamma\right)  }$. Thus, point $x_{1}^{-}$ is always a
source, $x_{1}^{+}$ is an attractor when $\lambda<\frac{1}{6\left(
\gamma-2\right)  ^{2}}$, while $x_{2}$ is the unique attractor when it exists.

\subsection{Ideal gas with $\Lambda\neq0$}

In the presence of the cosmological constant, that is, $\Lambda\neq0$, and
when the matter term is that of the ideal gas, the field equations are written
as follows%
\begin{equation}
\Omega_{m}=1-\lambda x^{2}-\Omega_{\Lambda}~, \label{ww.35}%
\end{equation}%
\begin{equation}
\Omega_{\Lambda}^{\prime}=-\Omega_{\Lambda}\left(  \left(  \gamma-2\right)
\lambda x^{2}+\gamma\left(  \Omega_{\Lambda}-1\right)  \right)  ~
\label{ww.36}%
\end{equation}
and
\begin{equation}
x^{\prime}=\frac{1}{12\lambda}\left(  \left(  \lambda x^{2}-1\right)  \left(
\sqrt{6}-6\left(  \gamma-2\right)  \lambda x\right)  +\left(  \sqrt{6}%
-6\gamma\lambda x\right)  \Omega_{\Lambda}\right)  ~. \label{ww.37}%
\end{equation}

Furthermore, we assume that $\left\vert \Omega_{\Lambda}\right\vert \leq1$,
from which we infer that $x$ is also bounded, and we do not have to study the
dynamical system for the existence of stationary points at infinity. \ 

The stationary points of the dynamics system (\ref{ww.36}), (\ref{ww.37}) are
defined in the plane $\left\{  x,\Omega_{\Lambda}\right\}  $, that is
$B=\left(  x\left(  B\right)  ,\Omega_{\Lambda}\left(  B\right)  \right)  $.
The points are
\begin{equation}
B_{1}^{\pm}=\left(  \pm\frac{1}{\sqrt{\lambda}},0\right)  ~,~B_{2}=\left(
\frac{1}{\sqrt{6}\left(  \gamma-2\right)  \lambda},0\right)  ~,
\end{equation}%
\begin{equation}
B_{3}=\left(  0,1\right)  ~,~B_{4}=\left(  \sqrt{6}\gamma,1+6\left(
2-\gamma\right)  \gamma\lambda\right)  \text{~}.
\end{equation}

Points $B_{1}^{\pm},~B_{2}\,$\ are actually the stationary points $A_{1}^{\pm
}$ and $A_{2}$ respectively for which the cosmological constant component is
zero. The physical properties are the same as before. However, we should
investigate the stability analysis.

For point $B_{3}$ we derive $w_{eff}\left(  B_{3}\right)  =-1$, $\Omega
_{m}\left(  B_{3}\right)  =0$. Thus point $B_{3}$ describes a de Sitter universe.

Furthermore, point $B_{4}$ provides $\Omega_{m}\left(  B_{4}\right)
=-12\gamma\lambda$, $w_{eff}\left(  B_{4}\right)  =-1$. The point is
physically acceptable when $-\frac{1}{24}\leq\lambda<0$, or $\lambda<-\frac
{1}{24}$ with $\gamma\leq-\frac{1}{12\lambda}$ or $\gamma=0$. The stationary
point describes the de Sitter universe in which all the fluid components
contribute in the cosmological solution.

We linearize the dynamical system (\ref{ww.36}), (\ref{ww.37}) around the
stationary points and we derive the eigenvalues. For points $B_{1}^{\pm}$ the
eigenvalues are $e_{1}\left(  B_{1}^{\pm}\right)  =2-\gamma\mp\frac{1}%
{\sqrt{6\lambda}}$, $e_{2}\left(  B_{1}^{\pm}\right)  =2$ from which we infer
that $B_{1}^{-}$ is always a source, while $B_{1}^{+}$ is a saddle point when
$\lambda<\frac{1}{6\left(  \gamma-2\right)  ^{2}}$. Otherwise it is a source.

For point $B_{2}$ the two eigenvalues are $e_{1}\left(  B_{2}\right)
=-1+\frac{\gamma}{2}+\frac{1}{12\lambda\left(  2-\gamma\right)  }$,
$e_{2}\left(  B_{2}\right)  =\gamma+\frac{1}{6\lambda\left(  2-\gamma\right)
}$. Thus, point is always a saddle point when it is physically acceptable
because $e_{1}\left(  B_{2}\right)  $ is always negative while $e_{2}\left(
B_{2}\right)  $ is always positive.

The eigenvalues of the linearized system around the de Sitter point $B_{3}$
are calculated to be $e_{1}\left(  B_{3}\right)  =-1$ and $e_{2}\left(
B_{3}\right)  =-\gamma$, from which we infer that the point is always an
attractor. Finally, for point $B_{4}$ we find the eigenvalues $e^{\pm}\left(
B_{4}\right)  =-\frac{1}{2}\pm\sqrt{1+4\gamma\left(  1+6\left(  2-\gamma
\right)  \lambda\right)  }$. Consequently, point $B_{4}$ is always a saddle point.

\subsection{Chaplygin gas with $\Lambda=0$}

Consider now that the matter source satisfies the equation of the state
parameter of a Chaplygin gas, $p_{m}=\frac{A_{0}}{\rho_{m}^{\alpha}}$, for
which $\alpha\geq1,~$ $A_{0}=\left(  -1\right)  ^{\alpha}3^{-\left(
1+\alpha\right)  }A$ and $\rho_{m}\neq0$. The field equations are written as
follows%
\begin{equation}
\Omega_{m}=1-\lambda x^{2}~, \label{ww.38}%
\end{equation}%
\begin{equation}
x^{\prime}=\frac{1}{12}\left(  \frac{\left(  \sqrt{6}+6\lambda x\right)
\left(  \lambda x^{2}-1\right)  }{\lambda}+6x\left(  \lambda x^{2}-1\right)
^{-\alpha}Y\right)  ~ \label{ww.39}%
\end{equation}
and
\begin{equation}
Y^{\prime}=\frac{1+\alpha}{6}Y\left(  6-\sqrt{6}x+6\lambda x^{2}+6\left(
\lambda x^{2}-1\right)  ^{-\alpha}Y\right)  ~, \label{ww.40}%
\end{equation}
where the new variable $Y$ is defined as $Y=Ae^{-\frac{1}{2}\left(
1+\alpha\right)  \phi}\theta^{-\left(  2+\alpha\right)  }$.

The stationary points $C=\left(  x\left(  C\right)  ,Y\left(  C\right)  \text{
}\right)  $of the dynamical system (\ref{ww.39}), (\ref{ww.40}), with
$\Omega_{m}>0$ are
\begin{equation}
C_{1}=\left(  -\frac{1}{\sqrt{6}\lambda},0\right)  ~,
\end{equation}%
\begin{equation}
C_{2}=\left(  \sqrt{\frac{3}{2}}-\frac{\sqrt{\lambda\left(  1+3\lambda\right)
}}{\sqrt{2}\lambda},\frac{\left(  \sqrt{3\lambda\left(  1+3\lambda\right)
}+6\lambda\left(  1+3\lambda-\sqrt{3\lambda\left(  1+3\lambda\right)
}\right)  \right)  \left(  -\frac{1}{2}-3\lambda\sqrt{3\lambda\left(
1+3\lambda\right)  }\right)  ^{\alpha}}{6\lambda}\right)  ~
\end{equation}
and
\begin{equation}
C_{3}=\left(  \sqrt{\frac{3}{2}}+\frac{\sqrt{\lambda\left(  1+3\lambda\right)
}}{\sqrt{2}\lambda},\frac{\left(  \sqrt{3\lambda\left(  1+3\lambda\right)
}+6\lambda\left(  1+3\lambda+\sqrt{3\lambda\left(  1+3\lambda\right)
}\right)  \right)  \left(  -\frac{1}{2}+3\lambda\sqrt{3\lambda\left(
1+3\lambda\right)  }\right)  ^{\alpha}}{6\lambda}\right)  ~.
\end{equation}

For point $C_{1}$ we derive $\Omega\left(  C_{1}\right)  =1-\frac{6}{\lambda}%
$, $w_{eff}\left(  C_{1}\right)  =\frac{1}{6\lambda}$. The point is physically
acceptable when $\lambda\geq\frac{1}{6}$ while it always describes a universe
without acceleration. For $\lambda=\frac{1}{6}$, the asymptotic solution is
that of dust, while for $\lambda=\frac{1}{2}$ the asymptotic solution is that
of radiation. The eigenvalues of the linearized system around the stationary
point are calculated $e_{1}\left(  C_{1}\right)  =\frac{\left(  1+\alpha
\right)  \left(  1+3\lambda\right)  }{3\lambda}$ ,~$e_{2}\left(  C_{1}\right)
=\frac{1-6\lambda}{12\lambda}$, from which we can easily conclude that the
stationary point is always a saddle point.

Point $C_{2}$ describes a universe for which $\Omega_{m}\left(  C_{2}\right)
=\frac{1}{2}-3\lambda+\sqrt{3\lambda\left(  1+3\lambda\right)  }$ and
$w_{eff}\left(  C_{2}\right)  =\lambda\left(  x\left(  C_{2}\right)  \right)
^{2}+\left(  \lambda\left(  x\left(  C_{2}\right)  \right)  ^{2}-1\right)
^{-\alpha}Y\left(  C_{2}\right)  $. The point is well defined when $\lambda
>0$, while for large values of $\lambda$ it follows that $w_{eff}\left(
C_{2};\lambda>>1\right)  \simeq-1$, which means that point $C_{2}$ can
describe a solution near to the de Sitter point. On the other hand, point
$C_{3}$ is physical acceptable for $0<\lambda\leq\frac{1}{24}$, while we
derive $\Omega_{m}=-3\lambda+\sqrt{3\lambda\left(  1+3\lambda\right)  }~$and
$w_{eff}\left(  C_{3}\right)  =\lambda\left(  x\left(  C_{3}\right)  \right)
^{2}+\left(  \lambda\left(  x\left(  C_{3}\right)  \right)  ^{2}-1\right)
^{-\alpha}Y\left(  C_{3}\right)  $ in which $w_{eff}\left(  C_{3}%
;\lambda=\frac{1}{24}\right)  =1$. Thus point $C_{3}$ does not describe any acceleration.

The eigenvalues of the linearized system near to the stationary points $C_{2}$
and $C_{3}$ are determined. Numerically we find that $e_{1}\left(
C_{2}\right)  ,~e_{2}\left(  C_{2}\right)  $ have always negative real parts
for $\lambda>0$ and $\alpha\geq1$; on the other hand $\operatorname{Re}\left(
e_{1}\left(  C_{3}\right)  \right)  \,>0,~\operatorname{Re}\left(
e_{2}\left(  C_{3}\right)  \right)  >0$ for $\alpha\geq1$ $,~0<\lambda
\leq\frac{1}{24}$. Hence, point $C_{2}$ is always an attractor while point
$C_{3}$ is always a source.

\subsection{Chaplygin gas with $\Lambda\neq0$}

In the presence of a nonzero cosmological constant term, the field equations
are reduced to the following dynamical system%
\begin{equation}
\Omega_{m}=1-\lambda x^{2}-\Omega_{\Lambda}~,
\end{equation}%
\begin{equation}
\Omega_{\Lambda}^{\prime}=\Omega_{\Lambda}\left(  1+\lambda x^{2}%
-\Omega_{\Lambda}+Y\left(  \lambda x^{2}+\Omega_{\Lambda}-1\right)  ^{-\alpha
}\right)  ~,
\end{equation}%
\begin{equation}
x^{\prime}=\frac{1}{12}\left(  x^{2}\left(  \sqrt{6}+6\lambda\right)
+\frac{\sqrt{6}}{\lambda}\left(  \Omega_{\Lambda}-1\right)  +6x\left(
Y\left(  \lambda x^{2}+\Omega_{\Lambda}-1\right)  ^{-\alpha}-1-\Omega
_{\Lambda}\right)  \right)  ~,
\end{equation}%
\begin{equation}
Y^{\prime}=\frac{1+\alpha}{6}Y\left(  6\left(  1+\lambda x^{2}+\Omega
_{\Lambda}+Y\left(  \lambda x^{2}+\Omega_{\Lambda}-1\right)  ^{-\alpha
}\right)  -\sqrt{6}\lambda x\right)  .
\end{equation}

The\ physically acceptable stationary points $D=\left(  x\left(  D\right)
,Y\left(  D\right)  ,\Omega_{\Lambda}\left(  D\right)  \right)  $ are%
\begin{equation}
D_{1}=\left(  x\left(  C_{1}\right)  ,Y\left(  C_{1}\right)  ,0\right)
~,~D_{2}=\left(  x\left(  C_{2}\right)  ,Y\left(  C_{2}\right)  ,0\right)  ~,~
\end{equation}%
\begin{equation}
D_{3}=\left(  x\left(  C_{3}\right)  ,Y\left(  C_{3}\right)  ,0\right)
~,~D_{4}=\left(  \sqrt{6},1+6\lambda,0\right)  ,
\end{equation}
where $D_{1},~D_{2}$ and $D_{3}$ have the same physical properties as points
$C_{1}$, $C_{2}$ and $C_{3}$ respectively.

For the point $D_{4}$ we find $\Omega_{m}\left(  D_{4}\right)  =-12\lambda$
and $w_{eff}\left(  D_{4}\right)  =-1$, which means that the asymptotic
solution is physically acceptable when $-\frac{1}{12}\leq\lambda<0$, while the
asymptotic solution is that of the de Sitter universe.

The eigenvalues of the linearized system near $D_{1}$ are $e_{1}\left(
D_{1}\right)  =\frac{\left(  1+\alpha\right)  \left(  1+3\lambda\right)
}{3\lambda}$ ,~$e_{2}\left(  D_{1}\right)  =\frac{1-6\lambda}{12\lambda}$ and
$e_{3}\left(  D_{1}\right)  =\frac{1+6\lambda}{6\lambda}$, which means that
point $D_{1}$ is always a saddle point. For the points $D_{2}$ and $D_{3}$ we
find numerically that $D_{2}$ is always an attractor while $D_{3}$ is always a
source. Finally, for the point $D_{4}$ we calculate $e_{1}\left(
D_{4}\right)  =-\left(  1+\alpha\right)  $,~$e_{2}^{\pm}=\frac{1}{2}\left(
-1\pm\sqrt{5+24\lambda}\right)  $, from which it follows that the stationary
point is always a saddle point.

\section{Minisuperspace description and conservation laws}

\label{sec4}

For an ideal gas $p_{m}=\left(  \gamma-1\right)  \rho_{m}$, from equation
(\ref{ww.20a}) it follows $\rho_{m}\left(  t\right)  =\rho_{m0}a^{-3\gamma
}e^{\phi}$ in which $\rho_{m0}$ is a constant of integration.

We substitute this into the rest of the field equations and we end with the
following dynamical system%
\begin{equation}
\frac{\theta^{2}}{3}-\frac{\lambda}{2}\dot{\phi}^{2}-\Lambda-\rho_{m0}%
e^{\frac{\phi}{2}}a^{-3\gamma}=0, \label{ee.01}%
\end{equation}%
\begin{equation}
\dot{\theta}+\frac{1}{3}\theta^{2}+\frac{\left(  3\gamma-2\right)  }{2}%
\rho_{m0}e^{\frac{\phi}{2}}a^{-3\gamma}+\lambda\dot{\phi}^{2}-\Lambda=0,
\label{ee.02}%
\end{equation}%
\begin{equation}
\ddot{\phi}+\theta\dot{\phi}+\frac{\rho_{m0}}{2\lambda}e^{\frac{\phi}{2}%
}a^{-3\gamma}=0. \label{ee.03}%
\end{equation}

For the second-order differential equations (\ref{ee.02}), (\ref{ee.03}) in
the space of variables $\left\{  a,\phi\right\}  $, the inverse problem for
the determination of a Lagrangian function, provides that the function
\begin{equation}
L\left(  a,\dot{a},\phi,\dot{\phi}\right)  =-3a\dot{a}^{2}+\frac{\lambda}%
{2}a^{3}\dot{\phi}^{2}-a^{3}\Lambda-\rho_{m0}e^{\frac{\phi}{2}}a^{3-3\gamma}
\label{ee.04}%
\end{equation}
is an autonomous Lagrangian function for the field equations, while equation
(\ref{ee.01}) is conservation law of \textquotedblleft
energy\textquotedblright, i.e. the Hamiltonian $\mathcal{H}$, constraint
$\mathcal{H}=0$.

In general, the field equations for the cosmological model in WIG theory with
an ideal gas, for the metric%
\begin{equation}
ds^{2}=-N^{2}\left(  t\right)  +a^{2}\left(  t\right)  \left(  dx^{2}%
+dy^{2}+dz^{2}\right)  , \label{ee.05}%
\end{equation}
follow from the singular point-like Lagrangian%
\begin{equation}
\mathcal{L}\left(  a,\dot{a},\phi,\dot{\phi}\right)  =\frac{1}{N}\left(
-3a\dot{a}^{2}+\frac{\lambda}{2}a^{3}\dot{\phi}^{2}\right)  -N\left(
a^{3}\Lambda+\rho_{m0}e^{\frac{\phi}{2}}a^{3-3\gamma}\right)  . \label{ee.06}%
\end{equation}

\subsection{Integrability property and analytic solution}

Because the field equations admit a point-like Lagrangian various techniques
inspired by analytic mechanics be applied for the study of the dynamical
system. Indeed, variational symmetries and conservation laws can be determined
by using Noether's theorems \cite{ns1}. That approach has been widely used in
various gravitational systems. New integrable cosmological models as also new
analytic and exact solutions were found through the use of variational
symmetries, see for instance \cite{ns2}.

We investigate for variational symmetries which have point transformations as
generators and provide conservation laws linear in the velocities. Hence, for
the Lagrangian function (\ref{ee.04}) and for $\rho_{m0}\neq0$, we find that
the variational symmetry $X=\frac{2}{3}a\partial_{a}+4\left(  \gamma-2\right)
\partial_{\phi}$ exists for $\Lambda=0$, and the corresponding conservation
law is
\begin{equation}
F\left(  a,\dot{a},\phi,\dot{\phi}\right)  =4a^{2}\dot{a}-4\left(
\gamma-2\right)  \lambda a^{3}\dot{\phi}-F_{0}. \label{ee.07}%
\end{equation}

Function $F\left(  a,\dot{a},\phi,\dot{\phi}\right)  $, $\frac{dF}{dt}=0$, is
the second-conservation law for the dynamical system, which means that the
field equations form an integrable dynamical system.

In order to reduce the field equations and determine exact solutions, we apply
the Hamilton-Jacobi approach. We define the momentum $p_{a}=-6a\dot{a}$, and
$p_{\phi}=\lambda a^{3}\dot{\phi}$., thus the Hamiltonian function
$\mathcal{H}\left(  a,\phi,p_{a},p_{\phi}\right)  =0$, reads%
\begin{equation}
-\frac{p_{a}^{2}}{6a}+\frac{p_{\phi}^{2}}{\lambda a^{3}}+2\left(  a^{3}%
\Lambda+\rho_{m0}e^{\frac{\phi}{2}}a^{3-3\gamma}\right)  =0 \label{ee.08}%
\end{equation}
while the Hamilton-Jacobi equation is written in the following form%
\begin{equation}
-\frac{1}{6a}\left(  \frac{\partial}{\partial a}S\left(  a,\phi\right)
\right)  ^{2}+\frac{1}{\lambda a^{3}}\left(  \frac{\partial}{\partial\phi
}S\left(  a,\phi\right)  \right)  ^{2}+2\rho_{m0}e^{\frac{\phi}{2}%
}a^{3-3\gamma}=0, \label{ee.09}%
\end{equation}
where now $p_{a}=\frac{\partial S}{\partial a}$ and $p_{\phi}=\frac{\partial
S}{\partial\phi}$.

Moreover, the conservation law (\ref{ee.07}) provides the constraint equation
for the Action $S\left(  a,\phi\right)  $%
\begin{equation}
\frac{2a}{3}\left(  \frac{\partial}{\partial a}S\left(  a,\phi\right)
\right)  +4\left(  \gamma-2\right)  \frac{\partial}{\partial\phi}\left(
S\left(  a,\phi\right)  \right)  -F_{0}=0. \label{ee.10}%
\end{equation}

We define the new variable $\phi=6\left(  \gamma-2\right)  \ln a+\Phi$, such
that the constraint equation becomes%
\begin{equation}
\frac{2}{3}a\frac{\partial}{\partial a}\left(  S\left(  a,\Phi\right)
\right)  -F_{0}=0. \label{ee.11}%
\end{equation}
This new set of variables $\left\{  a,\Phi\right\}  $ are the normal
coordinates for the dynamical system.

Consequently, in the normal variables the analytic expression for the Action
as provided by the Hamilton-Jacobi equation is
\begin{equation}
S\left(  a,\Phi\right)  =\frac{3}{2}F_{0}\ln a+\int\frac{\sqrt{2\lambda}%
\sqrt{16\rho_{m0}e^{\frac{\Phi}{2}}\left(  6\lambda\left(  \gamma-2\right)
^{2}-1\right)  +3F_{0}}+6\lambda\left(  \gamma-2\right)  F_{0}}{4\left(
6\lambda\left(  \gamma-2\right)  ^{2}-1\right)  }d\Phi\label{ee.12}%
\end{equation}
for $\left(  6\lambda\left(  \gamma-2\right)  ^{2}-1\right)  \neq0$, or%
\begin{equation}
S\left(  a,\Phi\right)  =\frac{3}{2}F_{0}\ln a+\frac{3F_{0}^{2}\Phi-32\rho
_{0}e^{\frac{\Phi}{2}}}{24F_{0}\left(  \gamma-2\right)  },~
\end{equation}
when $\left(  6\lambda\left(  \gamma-2\right)  ^{2}-1\right)  =0$.

However, in the new coordinates the momentum are defined as \
\begin{equation}
p_{a}=-6a\left(  \left(  6\lambda\left(  \gamma-2\right)  ^{2}-1\right)
\dot{a}+\left(  \gamma-2\right)  \lambda a\dot{\Phi}\right)  , \label{ee.13}%
\end{equation}%
\begin{equation}
p_{\Phi}=-\lambda a\left(  6\left(  \gamma-2\right)  \dot{a}+a\dot{\Phi
}\right)  , \label{ee.14}%
\end{equation}
which give the following expressions for the scale factor and the scalar
field
\begin{equation}
6a^{2}\dot{a}=ap_{a}-6\left(  \gamma-2\right)  p_{\Phi}, \label{ee.15}%
\end{equation}%
\begin{equation}
\lambda a^{3}\dot{\Phi}=-p_{\Phi}-\lambda\left(  \gamma-2\right)  \left(
Ap_{A}+6\left(  \gamma-2\right)  p_{\Phi}\right)  . \label{ee.16}%
\end{equation}

Hence, by using the Action (\ref{ee.12}) and expressions (\ref{ee.15}),
(\ref{ee.16}), the cosmological field equations can be written into an
equivalent system. We summarize the results in the following proposition.

\textbf{Proposition 1: }\textit{The field equations in WIG for a FLRW
background space with zero spatial curvatue and an ideal gas form a Liouville
integrable system when there is no cosmological constant term. The analytic
solution for the Hamilton-Jacobi equation provides the Action (\ref{ee.12}),
while the field equations can be written into an equivalent set of two
first-order ordinary differential equations (\ref{ee.15}), (\ref{ee.16}). }

Assume now the simple case for which $\gamma=1$ and $F_{0}=0$. Moreover, we
define the new variable $T=T\left(  t\right)  $, such that $dT=\frac
{\sqrt{\left(  6\lambda-1\right)  }}{A^{3}}dt$ and $\lambda\neq\frac{1}{6}$.

Thus, the field equations are
\begin{equation}
\frac{\dot{a}}{a}-\sqrt{2\lambda\rho_{m0}}e^{\frac{\Phi}{4}}=0,
\end{equation}%
\[
\dot{\Phi}-\sqrt{\frac{2}{\lambda}\rho_{m0}\left(  6\lambda+1\right)
}e^{\frac{\Phi}{4}}=0,
\]
with exact solution%
\begin{equation}
a\left(  t\right)  =a_{0}t^{\frac{4\lambda}{1+6\lambda}}~,~\Phi\left(
t\right)  =-2\ln\left(  \frac{\left(  6\lambda+1\right)  \rho_{m0}}{8\lambda
}t^{2}\right)  .
\end{equation}

For this exact solution the background space is%
\begin{equation}
ds^{2}=-\frac{\left(  6\lambda-1\right)  }{a_{0}^{6}}t^{-\frac{24\lambda
}{1+6\lambda}}dT^{2}+a_{0}^{2}t^{\frac{8\lambda}{1+6\lambda}}\left(
dx^{2}+dy^{2}+dz^{2}\right)  .
\end{equation}
The later solution describes a universe dominated by a perfect fluid source
with constant equation of state parameter. This specific solution is descibed
by the stationary points $A_{2}$, thus. the results are in agreement with the
asymptotic analysis for the dynamics.

\section{Conclusions}

\label{sec5}

In this work we considered WIG to describe the cosmological evolution for the
physical parameters in FLRW spacetime with zero spatially curvutare. The
gravitational field equations in WIG are of second-order and Einstein's
theory, with the presence of the of a scalar field, is recovered. Scalar field
plays the role for conformal factor which relates the connection of Weyl
theory with the Levi-Civita connection of Riemannian geometry. However, the
field equations differ when matter is introduced in the gravitational model.
Indeed, in WIG the matter source interacts with the scalar field. The
interaction term is introduced naturally from the geometric character of the theory.

In our study we considered the matter source to be described by that of an
ideal gas, that is $p_{m}=\left(  \gamma-1\right)  \rho_{m}$, or by the
Chaplygin gas $p_{m}=-\frac{A_{0}}{\rho_{m}^{\alpha}}$. We defined new
dimensionless variables based on the Hubble-normalization in order to write
the field equations as a system of first-order algebraic differential system.
In each model, we determined the stationary points for the latter system and
we determined their dynamical properties as also the physical properties of
the asymptotic solutions. In our analysis we considered also a nonzero
cosmological constant.

For the ideal gas, we found that there exists an attractor with an asymptotic
solution that of an ideal gas, but with different parameter for the equation
of state. For instance, we can consider the matter source to be that of
radiation while the attractor to describe an accelerated universe. In the
presence of the cosmological constant, we find two asymptotic solutions which
can describe the past acceleration phase of the universe known as inflation,
as also the late time acceleration. The future attractor describes the de
Sitter universe. When the matter component is that of a Chaplygin gas the
stationary points as also the cosmological evolution are similar with the
previous case.

Moreover, for the ideal gas case, we solved the inverse problem and determined
a Lagrangian function, and a minisuperspace description, which generates the
cosmological equations under a variation. We applied Noether's theorems for
point transformations in order to construct a nontrivial conservation law when
the cosmological constant term is zero. Hence, the cosmological field
equations form a Liouville integrable dynamical system. The closed-form
expression for the Hamilton-Jacobi equation derived. Finally, for a specific
values for the free parameters we were able to construct an exact solution
which is in agreement with the asymptotic analysis.

In a subsequent analysis we plan to investigate further the field equations as
a Hamilton system and understand how a nonzero cosmological constant affects
the integrability property of the field equations.


\begin{thebibliography}{99}                                                                                               %


\bibitem {pl1}Planck Collaboration: N. Aghanim et al. Planck 2018 results. VI.
Cosmological parameters, A\&A 641, A6 (2020)

\bibitem {leon1}L. Perivolaropoulos and F. Skara, Challenges for $\Lambda$CDM:
An update, (2021) [2105.05208]

\bibitem {sf1}B.\ Ratra and P.J.E. Peebles, Cosmological consequences of a
rolling homogeneous scalar field, Phys. Rev. D 37, 3406 (1988)

\bibitem {sf2}C. Armendariz-Picon, V.F. Mukhanov and P.J. Steinhardt, Phys.
Rev. D 63, 103510 (2001)

\bibitem {sf3}V. Faraoni, Cosmology in Scalar-Tensor Gravity, Kluwer Academic
Publishers, Dordrecht, (2004)

\bibitem {sf4}M. C. Bento, O. Bertolami and A. A. Sen, Generalized Chaplygin
gas and CMBR constraints, Phys. Rev. D 67, 063003 (2003).

\bibitem {sf5}A. Kamenshchik, U. Moschella and V. Pasquier, An alternative to
quintessence, Phys. Lett. B 511, 265 (2001)

\bibitem {sf5a}S. Basilakos, N.E. Mavromatos and J. Sol\`{a} Peracaula, Phys.
Rev. D 101, 045001 (2020)

\bibitem {sf6}T. Clifton, P.G.\ Ferreira, A. Padilla and C. Skordis,\ Modified
gravity and cosmology Phys. Rept., 513, 1, (2012)

\bibitem {sf7}S.I. Nojiri and S.D. Odintsov, Introduction to modified gravity
and gravitational alternative for dark energy, IJGMMP 4, 115 (2007)

\bibitem {sf8}E. Di\ Valentino, O. Mena, S. Pan, L. Visinelli, W. Yang, A.
Melchiorri, D.F. Mota, A.G. Riess and J. Silk, In the Realm of the Hubble
tension a Review of Solutions, Class. Quantum\ Grav. 38, 153001 (2021)

\bibitem {in0}A. P. Billyard and A. A. Coley, Interactions in scalar field
cosmology, Phys Rev D 61, 083503 (2000)

\bibitem {in2}S. Pan, G.S. Sharov and W. Yang, Field theoretic interpretations
of interacting dark energy scenarios and recent observations, Phys. Rev. D
101, 103533 (2020)

\bibitem {in4}W. Yang, A. Mukherjee, E. Di Valentino and S. Pan, Interacting
dark energy with time varying equation of state and the H0 tension,
Phys.\ Rev.\ D 98, 123527 (2018)

\bibitem {in5}S. Pan and G.S. Sharov, A model with interaction of dark
components and recent observational data, MNRAS 472, 4736 (2017)

\bibitem {in6}E. Di Valentino, A. Melchiorri, O. Mena and S. Vagnozzi,
Interacting dark energy in the early 2020s: a promising solution to the H0 and
cosmic shear tensions, Phys. Dark Univ. 30, 100666 (2020)

\bibitem {as1}S. Yu Vernov and E. Pozdeeva, De Sitter Solutions in
Einstein--Gauss--Bonnet Gravity, Universe 7, 149 (2021)

\bibitem {as2}A. Yu. Kamenshchik, E.O. Pozdeeva, G. Venturi and S. Yu Vernov,
Integrable cosmological models in the Einstein and in the Jordan frames and
Bianchi-I cosmology, Phys. Part. Nucl. 49, 1 (2018)

\bibitem {as3}N. Dimakis, A. Paliathanasis, P.A.\ Terzis and T.
Christodoulakis, Cosmological solutions in multiscalar field theory, EPJC 79,
618 (2019)

\bibitem {as4}A. Paliathanasis, De Sitter and scaling solutions in a
higher-order modified teleparallel theory, JCAP 08, 027 (2017)

\bibitem {dn1}E.J. Copeland, A.R. Liddle and D. Wands, Exponential potentials
and cosmological scaling solution, Phys. Rev. D 57, 4686 (1998)

\bibitem {dn2}A. Coley and G. Leon, Static Spherically Symmetric
Einstein-aether models I: Perfect fluids with a linear equation of state and
scalar fields with an exponential self-interacting potential, Gen. Rel. Grav.
51, 115 (2019)

\bibitem {dn3}L. Amendola, D. Polarski and S. Tsujikawa, Are f(R) dark energy
models cosmologically viable ?, Phys. Rev. Lett. 98, 131302 (2007)

\bibitem {dn4}L. Amendola, R. Gannouji, D. Polarski and S. Tsujikawa,
Conditions for the cosmological viability of f(R) dark energy models,
Phys.\ Rev. D\ 75, 083504 (2007)

\bibitem {dn5}P. Christodoulidis, D. Roest and E.I. Sfakianakis, Scaling
attractors in multi-field inflation, JCAP 12, 059 (2019)

\bibitem {dn6}C.R. Fadragas, R. Cardenas, M. Rodriguez-Ricard, A.
Rivero-Acosta and A. Linares-Rodriguez, Detailed qualitative dynamical
analysis of a cosmological Higgs field, Gen. Rel. Gravit. 51, 109 (2019)

\bibitem {dn7}T. Gonzalez, G. Leon and I. Quiros, Dynamics of quintessence
models of dark energy with exponential coupling to dark matter, Class. Quantum
Grav. 23, 32165 (2006)

\bibitem {dyn8}M. Kerachia, G. Acquaviva and G. Lukes-Gerakopoulos, Dynamics
of classes of barotropic fluids in spatially curved FRW spacetimes,
Phys.\ Rev. D\ 101, 043535 (2020)

\bibitem {sc1}J.E. Madriz Aguilar and C. Romero, Inducing the cosmological
constant from five-dimensional Weyl space, Found. Phys. 39, 1205 (2009)

\bibitem {sc2}Y.-X. Liu, K. Yang and Y. Zhong, de Sitter Thick Brane Solution
in Weyl Geometry, JHEP 10, 069 (2010)

\bibitem {sc3}I.P. Lobo, A.B. Barreto and C. Romero, Space-time singularities
in Weyl manifolds, EPJC 75, 448 (2015)

\bibitem {sc4}A. Paliathanasis and G. Leon, Integrability and cosmological
solutions in Einstein-\ae ther-Weyl theory, EPJC 81, 255 (2021)

\bibitem {sc5}M.L. Pucheu, F.A.P. Alves Junior, A.B. Barreto and C. Romero,
Cosmological models in Weyl geometrical scalar-tensor theory, Phys.\ Rev. D
94, 064010 (2016)

\bibitem {sc6}J. Miritzis, Isotropic cosmologies in Weyl geometry, Class.
Quantum Grav. 21, 3043 (2004)

\bibitem {sc7}J.M. Salim and S.L. Saut\'{u}, Gravitational collapse in Weyl
integrable space-times, Class. Quantum Grav. 16, 3281 (1999)

\bibitem {salim96}J.M. Salim and S.L. Saut\'{u}, Gravitational theory in Weyl
integrable spacetime, Class. Quantum Grav. 13, 353 (1996)

\bibitem {ns1}A. Halder, A. Paliathanasis and P.G.L.\ Leach, Noether's Theorem
and Symmetry, Symmetry 10, 744 (2018)

\bibitem {ns2}M. Tsampalis and A. Paliathanasis, Symmetries of Differential
Equations in Cosmology, Symmetry 10, 233 (2018)
\end{thebibliography}
\end{document}